\documentclass{article}
\usepackage{spconf,amsmath,graphicx}
\usepackage{amssymb}

\usepackage{algorithm}
\usepackage[noend]{algpseudocode}

\usepackage[table,xcdraw]{xcolor}
\usepackage{adjustbox}

\makeatletter
\def\BState{\State\hskip-\ALG@thistlm}
\makeatother

\usepackage{url}
\usepackage{cite}
\usepackage{booktabs}
\usepackage{color}
\usepackage{microtype}
\usepackage{multirow}

\DeclareMathOperator*{\argmin}{argmin}
\DeclareMathOperator*{\median}{median}

\title{Shift-Invariant Kernel Additive Modelling for Audio Source Separation}
\name{Delia Fano Yela$^1$, Sebastian Ewert$^{2}$, Ken O'Hanlon$^1$ and Mark B. Sandler$^1$ \thanks{This work was funded by EPSRC grant EP/L019981/1 and was conducted while S.~Ewert was at Queen Mary University of London. The authors would like to thank Giulio Moro, Gijs Wijnholds and Adib Mehrabi.}}
\address{$^1$ Queen Mary University of London, UK  \;\;\; $^2$ Spotify}

\begin{document}
\ninept

\maketitle
\begin{abstract}
A major goal in blind source separation to identify and separate sources is to model their inherent characteristics. While most state-of-the-art approaches are supervised methods trained on large datasets, interest in non-data-driven approaches such as Kernel Additive Modelling (KAM) remains high due to their interpretability and adaptability. KAM performs the separation of a given source applying robust statistics on the time-frequency bins selected by a source-specific kernel function, commonly the K-NN function. This choice assumes that the source of interest repeats in both time and frequency. In practice, this assumption does not always hold. Therefore, we introduce a shift-invariant kernel function capable of identifying similar spectral content even under frequency shifts. This way, we can considerably increase the amount of suitable sound material available to the robust statistics. While this leads to an increase in separation performance, a basic formulation, however, is computationally expensive. Therefore, we additionally present acceleration techniques that lower the overall computational complexity.
\end{abstract}
%

\section{Introduction}
\label{sec:intro}

Music recordings are typically produced by mixing a large number of instrument tracks, corresponding to vocals, guitars, drums or one of various synthesizers. This process makes analysing and processing music highly challenging as the individual instruments are usually strongly correlated in both time and frequency. Spatial information contained in the two stereo channels is often unreliable due to the use of various  non-linear sound effects yielding artificial sound scenes which cannot physically be reproduced and are difficult to model. 
Given such constraints, a major goal is to find inherent characteristics of the sources to identify and extract a target. Examples include the temporal behaviour of an instrument (continuity in time \cite{Virtanen07_MonauralSoundSourceSeparation_TASLP}, vibrato structures \cite{RegnierP09_SingingVoiceVibratoDetection_ICASSP}) or its spectral characteristics (spectral envelope\cite{DurrieuRDF10_SourceFilterMelody_IEEETASL}, percussive versus harmonic properties \cite{OnoMLKS08_ComplementaryDiffucion_ESPC}). 

Most state-of-the-art methods are based on either Non-Negative Matrix Factorisation (NMF) \cite{SchmidtM06_ShiftInvNMF_ICABSS,SmaragdisRS08_ShiftPLCA_ICASSP,OzerovVB12_PriorInfoSourceSep_TASLP} or Deep Networks \cite{UhlichGM15_DNNSeparation_ICASSP}, with each having different trade-offs with respect to run-time, separation quality and adaptability to new acoustic conditions. Both approaches are widely used in settings where (large amounts of) training material is available, which enables supervised learning and typically yields performance improvements. However, despite a measurable  difference in performance, interest in non-data-driven methods remains high: a focus on modelling concepts explicitly often increases the interpretability of methods, which opens more angles for including prior knowledge, might help with understanding how data-driven methods operate and can lead to a high generalization capacity across datasets.

In this context, Kernel Additive Modelling (KAM) has been successfully employed for a variety of tasks in source separation, such as vocal separation, speech enhancement or interference reduction  \cite{LiutkusFRPD14_KernelAdditive_IEEE-TSP,RafiiP13_RepetSimSpeech_ICASSP,FanoYelaEFS17_HybridKamNmf_ICASSP}. The core idea is related to Gaussian processes (GPs): Given a time frequency representation, one makes the assumption that individual entries correlate with others in a known way -- in other words, if we can observe the value of one entry, we can make a statement about the value of the related entries. An important difference between KAM and GPs is that in the latter an estimate is obtained as a solution of an inference problem, which involves feedback between values and thus is relatively slow. In KAM, feedback does not exist in this form, which can limit its expressivity on the one hand but allows for non-Gaussian relationships, which in practice enables the use of outlier resistant methods from robust statistics, and also leads to a drastic improvement in terms of computational performance. 

A central goal in KAM is to design a function (or \emph{kernel}) that, given a bin in a time-frequency representation, identifies bins having a similar contribution from a given source, ignoring the entries associated with other sources. If the magnitude of a bin deviates from the remaining ones defined by the kernel, one can assume that another source is present in that bin and that the bins in the kernel can be used to reconstruct the original value in the overlaid bin. KAM employs order statistics to attenuate the influence of outliers originating from other sources. 

A popular kernel choice in KAM is the K nearest neighbours (K-NN) function finding the most similar time frames, based on the squared Euclidean distance \cite{LiutkusFRPD14_KernelAdditive_IEEE-TSP}. This simple kernel function implicitly relies on two assumptions. Firstly, it assumes the energy in each time frame to be dominated by the source of interest (in \cite{FanoYelaEFS17_HybridKamNmf_ICASSP} the authors present KAM extensions for low SNR conditions). Secondly, using the Euclidean distance between entire frames, the position of partials and other objects cannot change. In other words, frames are required to repeat with only minor modifications. While this might be a valid assumption for full-length pop songs, it might be wrong if the recording is short, the source is consistently overlaid with the same interference in each repetition or for sources with highly variable pitch.

In this paper, we propose an extension to the KAM framework in the form of a shift-invariant kernel to overcome these limitations.
In particular, using a logarithmic frequency axis, our kernel extends the K-NN function by comparing not only the original frames but also all shifted versions. In other words, it can identify notes of the same source differing in pitch as similar and reconstruct a unique musical event from them
despite the shift. This way, our method drastically increases the sound material available for the sound reconstruction.
In a basic version our shift invariant extension is computationally quite expensive as distances have to be computed for various shifts. Therefore, we present a technique to lower the computational complexity and runtime of our proposed kernel considerably: taking inspiration from \cite{SaitoKTNS08_specmurtTranscription_TASLP}, instead of computing all shifts, we transform our logarithmic time-frequency representation into the magnitude specmurt domain, which enables an efficient comparison of frames based on their repetitive structures in frequency direction while ignoring the exact location of those structures. This way, we can construct a highly efficient method yielding a pre-selection of frames, which can then easily be pruned. 

The paper is structured as follows. In Section \ref{sec:pmethod}, we describe the baseline version of KAM and our proposed extensions. In Section \ref{tab:eval}, we apply our proposed method in a studio recording scenario, where the task is to restore short clips of individual instruments and remove interferences such as coughs or door slams. We conclude in Section \ref{sec:conclusion} with an outlook on future work.

\section{Proposed Method}
\label{sec:pmethod}

\subsection{KAM Baseline}

KAM is a rich framework with a wide range of applications \cite{LiutkusFRPD14_KernelAdditive_IEEE-TSP}. In the following, we limit the description of our baseline approach to the necessary level -- our extensions, however, are just as valid in the full framework.
In particular, consider a mixture $x(t) = s(t) + n(t)$ of two different sound sources $s$ and $n$. We assume $s$ to be energetically dominant in the mixture and the support of $n$ to be known and be limited to a short duration. For scenarios in which these two assumptions are not met, we refer to \cite{FanoYelaEFS17_HybridKamNmf_ICASSP,FanoYelaEFS17_TemporalContext_AES} for a set of additional KAM extensions. The task is to recover $s$ from the given mixture $x$ when $n$ is active.

\begin{figure}[t!]
\centering
\includegraphics[width= 6cm,height=10cm]{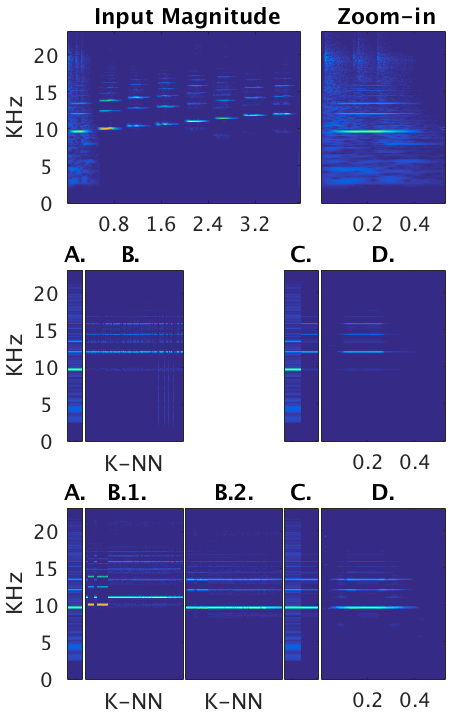}
\caption{Comparison of the baseline (second row) and the basic version of our proposed shift-invariant method (third row) for an example frame (A) from the input magnitude frames overlaid by an interference (Zoom-in). The K closest frames found for the current frame (A) by the baseline (B) and the proposed method, before (B.1) and after the shifting operation (B.2). The plots (C) contain the current frame next to the estimated output frame for each method. The complete estimation of the harmonic source for the frames containing interference is shown in D for both methods.}
\label{fig:methodIllustration}
\end{figure}

In the following, let $X, S \in \mathbb{C}^{F \times T}$ be time-frequency representations of $x$ and $s$, respectively, and $\overline{X}, \overline{S}$ the corresponding magnitudes. For KAM, we define a similarity kernel function $\mathcal{I}: F\times T \rightarrow \mathcal{P}(F\times T)$  that assigns to every time-frequency bin $(f,t)$ in $S$ a list of $K$ bins to be called similar (i.e.\,$\forall (f,t) \in F\times T: |\mathcal{I}(f,t)| = K$). In the following, the kernel function is the $K$-nearest neighbours ($K$-NN) function based on the squared Euclidean distance. In particular, for every time-frequency bin $(f,t)$, the bin $(f,\tilde{t})$ will be in $\mathcal{I}(f,t)$ if the time frame $\tilde{t}$ is among the $K$ most similar time frames. 

With $\mathcal{I}(f,t)$ defined, we know which bins are similar in $S$. If a bin in $S$ is overlaid by energy corresponding to $N$, we can use the similar bins in the observed $X$ to identify their commonalities and restore the overlaid bin. To this end, we express this estimation problem in KAM as a minimization of a \emph{model cost function} $\mathcal{L}$, which can be stated for a single channel as follows:
\begin{equation}
\overline{S}(f,t) \approx \argmin_{\lambda \in \mathbb{R}} \sum_{ (f,\tilde{t}) \in \mathcal{I}(f,t)}{\mathcal{L}(\overline{X}(f,\tilde{t}),\lambda)}.
\label{eq:KAMModelCostProblem}
\end{equation}
Depending on the choice of $\mathcal{L}$ the information in the bins indexed by $\mathcal{I}$ is merged in different ways.
The choice should take into account that, while all these bins are similar in $S$, there might be considerable, non-Gaussian differences between them in $X$ due to the unknown interference $n$. A popular choice is
$\mathcal{L}(a,b) := |a-b|$ as it leads to solutions employing operators from robust statistics (order statistics), which enable unbiased parameter estimation in the presence of up to 50\% outliers. With this choice of $\mathcal{L}$, the solution of the estimation problem~(\ref{eq:KAMModelCostProblem}) is:
\begin{equation}
\overline{\mathcal{S}}(f,t) := \median( \overline{X}({f},\tilde{t})  | ({f},\tilde{t}) \in \mathcal{I}(f,t)).
\label{eq:MedianIsSolution}
\end{equation}
$\overline{\mathcal{S}}$ being the magnitude estimate of the source of interest $s$, we define a corresponding magnitude estimate for the remaining sources in the mixture $n$ as $\overline{\mathcal{N}} = \max(\overline{X} - \overline{\mathcal{S}},0)$. Then we can perform the actual separation through soft masking and obtain a complex estimate $\mathcal{S}$ of the source of interest via $\mathcal{S} = \frac{\overline{\mathcal{S}}}{\overline{\mathcal{N}} + \overline{\mathcal{S}}} \odot X$, which is converted to the time-domain using an inverse time-frequency transform. 

The success of the separation heavily depends on the ability of the kernel to identify similar frames in the presence of overlaying sources. 
Using just a Euclidean distance between entire frames, this notion of similarity, however, can be quite limited. For example, as it can be seen in the second row of Fig.\ref{fig:methodIllustration}, the method might not be able to remove an interference on top of a single note played only once, Fig\ref{fig:methodIllustration}.A (as there might be no other similar frames). In particular, the method can not make use of frames where the instrument plays also a note but in a different pitch -- due to the difference in pitch the two frames are likely to be orthogonal, which leads to high differences in the Euclidean distance, as shown by the selection of $K$-NN in Fig\ref{fig:methodIllustration}.B. 

For our complexity analysis below, note that taking $X \in \mathbb{C}^{F \times T}$ as the input of our system (usually with $T>F$), the overall complexity of this baseline method is $O(T^2 (F + \log T))$. 

\subsection{Use of the Log-Frequency Domain}
KAM implementations typically use a standard linear scale time-frequency representation as it is both memory and computationally inexpensive. In such a representation, the spacing between harmonics and fundamental frequency will depend on the latter. However, using a logarithmic frequency scale the location of every harmonic with respect to the fundamental frequency will be constant \cite{Brown91_ConstantQ_JASA}. More precisely, taking $f_{0}$ as the fundamental frequency of a signal, the frequency of the $n^{th}$ harmonic will be located at $n \times f_{0}$ in a linear scale but would appear at $ \log f_{0} + \log n$ in a logarithmic frequency scale \cite{SchorkhuberKHD14_CQTToolbox_AES}. In particular, within a certain frequency range, pitch shifts simply correspond to shifts in log-frequency representations. 

\subsection{Shift-Invariant KAM}
\label{sec:SIKAM}

For our extension to the baseline kernel we make use of this property. In particular, let $X, S \in \mathbb{C}^{F \times T}$ be the Constant-Q transform (CQT) of $x$ and $s$, a log-frequency representation with a perfect reconstruction property \cite{SchorkhuberKHD14_CQTToolbox_AES}.
The goal now is to locate not only patterns repeated in time but also their shifted versions. In order to do so, we introduce a shift $ \delta$ in the kernel function measured in frequency bins. 
To this end, let $X_\delta$ be a frequency shifted version of X: $X_\delta(f,t) := X(f+\delta,t)$. We define our new shift-invariant kernel 
$\mathcal{I}_s$ as follows: For a given $(f,t)$, we have $(\tilde{f},\tilde{t}) \in \mathcal{I}_s(f,t)$ if $|\delta|<\Delta$ for $\delta := \tilde{f} - f$ and $\overline{X}_\delta(:,\tilde{t})$ is among the $K$ closest frames for frame $\overline{X}(:,t)$ across all $\delta \in \{-\Delta,\ldots,\Delta\}$. Here, we used the slicing notation $:$ to denote all elements in an index dimension. This means that two time frames can now be considered as neighbours if they display a similar harmonic pattern at different frequency locations. In other words, the proposed kernel function $\mathcal{I}_s$ can be seen as a shift-invariant version of the baseline kernel $\mathcal{I}$.  The estimation problem remains essentially the same (just that the variability in frequency is now explicit):
 \[
\overline{S}(f,t) \approx \argmin_{\lambda \in \mathbb{R}} \sum_{ (\tilde{f},\tilde{t}) \in \mathcal{I}_s(f,t)}{\mathcal{L}(\overline{X}(\tilde{f},\tilde{t}),\lambda)}.
\]
For the same model cost function as above, we get the solution:
\[
\overline{\mathcal{S}}(f,t) := \median( \overline{X}(\tilde{f},\tilde{t})  | (\tilde{f},\tilde{t}) \in \mathcal{I}_s(f,t)).
\]
As a result of this extension, we can now recover a note played only once by using notes different in pitch played by the same instrument, as seen in the third row of Fig\ref{fig:methodIllustration}.

In practice, the implementation of this approach can be split into two main steps: similarity measure (Fig\ref{fig:methodIllustration}.B.1) and frequency alignment (Fig\ref{fig:methodIllustration}.B.2). 
In particular, every frame in the mixture has to be shifted in frequency direction and compared to the remaining frames, $2 \cdot \Delta$ times. Computing the Euclidean distances in every step is in $O(T^2\cdot F)$. 
All together, with $\Delta$ typically being dependent on $F$, the complexity of this approach is considerable: $O(T^2 ( F^2 + \log T ))$. In practice, even after limiting $\Delta$ to a reasonable frequency range (such as 1-2 octaves), a basic implementation of this approach turns out to be computationally quite expensive.

\subsection{Acceleration Extension}
Under runtime constraints, the method above forces the user to trade-off separation performance for better running time. For example, one may set the $\Delta$ to cover only half an octave, at the risk of not finding similar events. In this section, we describe techniques to accelerate the kernel defined in Section~\ref{sec:SIKAM} considerably, while preserving the increase in separation quality.

\subsubsection{Similarity measure}
\label{sec:accelSimMeas}

Instead of applying the kernel function on the magnitude of the CQT, we propose to use a different time-frequency representation to allow a quicker shift-invariant search. The idea it to employ a representation that captures the (harmonic) patterns in each frame, while being invariant against their exact location. More precisely, given the magnitude CQT $\overline{X}$, we perform our search on the magnitude spectrum calculated on each frame $\overline{X}(:,t)$. This transform is related to cepstral analysis \cite{RabinerJ93_fundamentalsSpeech_BOOK} but has more recently been called \emph{specmurt} analysis~\cite{SaitoKTNS08_specmurtTranscription_TASLP} when applied to a log-frequency linear-magnitude representation (as in our case). 

Using the specmurt domain brings various advantages. First, eliminating the specmurt-phase, we eliminate pitch information and keep only the 'pattern' information. Second, certain spectral characteristics are represented more compactly. For example, a broadband sound in the time-frequency domain will correspond to 'low-frequency' components in the specmurt domain. This way, percussive components can more easily be ignored in the similarity search (if needed) and provides an interesting new angle to design source specific kernels by applying different weightings to the specmurt coefficients. 
Further, we can exploit the symmetry of the Fourier transform to eliminate half of the specmurt components, reducing the run time further.

Overall, instead of $O(T^2 ( F^2 + \log T ))$ operations for the shifts and Euclidean distances as before, we transform $\overline{X}$ to specmurt and only have to perform one set of Euclidean distances (comparable to the baseline that does not support shift invariance), resulting in only $O(T^2 ( F + \log T))$ operations for these steps.

\subsubsection{Frequency Alignment}

While the approach in Section~\ref{sec:accelSimMeas} enables a rapid shift-invariant selection of frames, it does not provide the shift we need to apply to a frame such that it is indeed similar to a given one. Given an input frame, a first idea is to apply all possible shifts to the $K$ frames found as similar on the specmurt domain. While this is a considerable speedup over the plain approach described in Section~\ref{sec:SIKAM}, it is still rather slow. Therefore, we will accelerate this step next, again using the Fourier transform, which was explored in \cite{SaitoKTNS08_specmurtTranscription_TASLP} in a related form in the context of source-filter modelling.

To this end, we assume we know from the last step that frames $t$ and $\tilde{t}$ are similar. For notational purposes, we will use the shorthands $Y := \tilde{X}(:,t)$ and $Z := \tilde{X}(:,\tilde{t})$. That means, $Y$ and $Z$ differ mostly by a shift in frequency, which we need to identify. We can express this as $Y = H * Z$ and solve for $H$ -- in case $Y=Z$, $H(0)=1$ and there is no shift. If all entries in $Z$ are shifted by $1$ compared to $Y$, we obtain $H(1)=1$. That means, to obtain the correct shift between $Y$ and $Z$ we only need to compute a deconvolution between them -- and the Fourier transform can again accelerate this step. As detailed in \cite{SaitoKTNS08_specmurtTranscription_TASLP}, a fast deconvolution can be calculated via
\begin{equation}
H = \mathcal{F} \left( \dfrac {\mathcal{IF}(Y)}{ \mathcal{IF}(Z)} \right),
\end{equation}
where $\mathcal{IF}$ and $\mathcal{F}$ denote the (inverse) Fourier transform.

Assuming that frames $t$ and $\tilde{t}$ are indeed similar, the $H$ we obtain this way, will typically be very sparse and essentially have a strong peak at exactly one position, which indicates the shift we need to apply to frame $\tilde{t}$. Once we have the optimal shift for all $K$ close frames we can continue as in the baseline method. Combining the two acceleration methods, the computational complexity is $O(T^2 ( F + \log T))+ O(T \cdot F \log F)$.

\subsubsection{ Pruning}

Even though measuring similarity based on the magnitude specmurt considerably reduces the computational complexity, it does not assure the frames found to be similar are the most similar. Discarding the phase in the kernel function renders the method shift-invariant but it also eliminates the unitary property of the Fourier Transform, i.e. Parseval's theorem does not hold anymore and thus Euclidean distances can be different. Therefore, when measuring the Euclidean distance between two frames in the magnitude specmurt, a large distance certainly indicates dissimilarity but a small distance does not assure a close match in the time-frequency domain (for example, major and minor chords can get confused).

To overcome this drawback while maintaining the complexity reduction, we here propose to use our acceleration technique as a pruning method. Instead of selecting K-NN in the kernel function, we select a larger fixed value $(K+P)$ to increase the pool of close frames. We then perform the specmurt analysis described above to find their optimal shift. At this point, one can retrieve these $(K+P)$ frames in the time-frequency representation and shift them by their corresponding amount. This means, we now have a narrowed down shifted version of the input magnitude, and so we can apply the baseline method to select the K-NN from the $(K+P)$ frames presented.
The overall complexity remains the same.

\section{Evaluation}
\label{tab:eval}
We evaluated the proposed method for an interference reduction application, where a burst-like sound overlays the audio recording. In particular, we focused on four different interferences that typically occur in live or studio recording scenarios: cough, chair drag sound, door slam and sound of object being dropped. We retrieved  example recordings of each from \url{freesound.org}.

In the following, we are mainly interesting in finding out how the different methods behave on recordings where the musical source is not repeated in time.
To this end, we created a synthetic dataset where the repeated and not repeated passages are known, so that we were able to compare the proposed method against the baseline in both cases. We created five different melodies (monophonic) and five different chord progressions, to simulate short studio takes, and synthesized these with 12 different instruments using the high quality Native Instruments Komplete Ultimate suite. We then created test recordings by overlaying the recordings with the interferences at 12 dB SNR, placing the interferences at two different locations: on a repeated musical segment and on a not repeated one, resulting on 960 tracks between 5 and 10 seconds each. While a more realistic dataset might better indicate the performance of the methods, we chose this setup to investigate exactly those cases where the individual methods might differ the most.

To quantitatively compare the separation quality of our proposed extension to the baseline, we used the BSS Eval toolbox 3.0 \cite{VincentGF06_PerformanceMeasurement_IEEE-TASLP} to calculate the Signal to Distortion Ratio (SDR). We used the CQT implementation described in \cite{SchorkhuberKHD14_CQTToolbox_AES}, setting the parameters to 24 bins per octave, gamma value of 20, minimum frequency of 27.5Hz and the maximum frequency being half of the sampling frequency (44.1KHz). For all methods, we set the parameter $K$ of the $K$-NN kernel function to 300 frames. Note however, that $K$ can and should be adjusted to the level of repetitiveness in the input recordings -- the higher the repetitiveness, the more all methods benefit from higher $K$.
For our proposed method, we fixed the number of shifts $\Delta$ to 48 (covering 4 octaves in total). In the acceleration+pruning method, the parameter $P$ of the $(K+P)-NN$ kernel function is set to be $2K$. We ignored the first coefficient in the specmurt representation as we expect it to mainly capture the broadband components. In addition, we assume the location of the interference in the mixture is known (and refer to \cite{FanoYelaEFS17_HybridKamNmf_ICASSP} otherwise) and thus we only process the frames affected and measure the SDR on those segments. The kernel function for all methods is applied to the remainder of the frames. 

\begin{table}
\scalebox{0.95}{
\centering
{\footnotesize
\begin{tabular} {c c c c c c c }\toprule
& \multicolumn{2}{c}{Melody} & \phantom{abc}& \multicolumn{2}{c}{Chords} \\
\cmidrule{2-3} \cmidrule{5-6} 
		& Repeated & Not repeated && Repeated & Not repeated  \\ \midrule
Baseline 	&  3.31	&  -2.40 			&& 4.11 			& 1.26  \\
Prop. 01 	&  4.61 & 3.87 				&& 4.11 			& $\mathbf{2.11}$  \\
Prop. 02	&  5.06 & 4.22				&& 4.03 			& 1.09\\
Prop. 03	&   $\mathbf{5.23}$ &  $\mathbf{4.36}$ 				&& $\mathbf{4.52}$ 	& 2.10\\

\bottomrule
\end{tabular}}
}
\caption{NSDR values for the baseline, the basic shift-invariant proposed method (Prop. 01) and the acceleration technique without pruning (Prop.02) and with pruning from an initial pool of twice the amount of K frames (Prop. 03). Parameters: K = 300, SNR = 12dB, $\Delta$  = 48}
\label{tab:results}
\vspace{-0.5cm}
\end{table}

The results with respect to the normalised SDR (NSDR) are given in Table~\ref{tab:results}, for both melody and chord progressions, on repeated and not repeated musical segments. As expected, the KAM baseline behaves poorly when there is no repetition, especially for melodies, which resembles the common scenario in popular songs where the source of interest is consistently repeating on the same pattern of unwanted sources. The NSDR value for the baseline for the non-repeated chords shows that, even though the chord is not repeated, some of its notes might, which can already be exploited by the method. However, the basic shift-invariant method ($Prop.~01$) clearly outperforms the baseline in those not repeated cases demonstrating standard KAM's limitations in such cases. In addition, it matches or improves the performance of the baseline on repeated segments, which suggests the proposed kernel function benefits from the shifting operation presenting the overlaying unwanted sources as clear outliers (affected by a shift in frequency). The basic shift-invariant method remains computationally expensive. However, the proposed methods based on specmurt analysis with ($Prop.~03$) and without pruning ($Prop.~02$) are effective in the melody scenario by even improving upon $Prop.~01$'s separation performance. This can be explained by the fact that the accelerated variants can find arbitrary shifts, while the shift in $Prop.~01$ is limited to reduce the computational time. In the chord progressions scenario, the low results of $Prop.~02$ confirms limitations in using the specmurt domain and justifies its use as a pre-selector for the pruning method $Prop. 03$.

\section{Conclusion}
\label{sec:conclusion}

We have presented an extension to the KAM framework in the form of a shift-invariant kernel function, aiming to overcome KAM's limitations with respect to non-repeating musical passages of the source of interest. We introduced a frequency shift in the kernel function to incorporate instances of the source of interest with similar frequency pattern at different frequency locations, increasing the pool of similar frames available for the source's reconstruction. Firstly we described a basic implementation bearing a high computational complexity and then presented acceleration techniques, which considerably lowered the computational complexity and runtime. The proposed methods were evaluated in an interference reduction scenario for transient noise typically found on live and studio music recordings. The results clearly demonstrate the inability of the baseline kernel to reconstruct non-repeated musical events and confirms the efficacy of the proposed shift-invariant kernel for such cases. However, even for repeated segments, the increase of the pool of similar frames led to improvements over standard KAM. Possible future directions for extending this work include an implementation for naturally sparse sources such as vocals.

\vfill\pagebreak
\bibliographystyle{bibstyle}
\bibliography{refsnew,referencesMusic}

\end{document}